\documentclass[fleqn,10pt]{wlscirep}
\usepackage[utf8]{inputenc}
\usepackage[T1]{fontenc}
\title{Environment driven oscillation in an off-lattice May--Leonard model}

\author[1]{D. Bazeia}
\author[2]{M. J. B. Ferreira}
\author[2]{B.F. de Oliveira}
\author[3]{A. Szolnoki}
\affil[1]{Departamento de Física, Universidade Federal da Paraíba, 58051-970 João Pessoa, PB, Brazil}
\affil[2]{Departamento de Física, Universidade Estadual de Maringá, Av. Colombo 5790, 87020-900 Maringá, PR, Brazil}
\affil[3]{Institute of Technical Physics and Materials Science, Centre for Energy Research, P.O. Box 49, H-1525 Budapest, Hungary}

\begin{abstract}
Cyclic dominance of competing species is an intensively used working hypothesis to explain biodiversity in certain living systems, where the evolutionary selection principle would dictate a single victor otherwise. Technically the May--Leonard models offer a mathematical framework to describe the mentioned non-transitive interaction of competing species when individual movement is also considered in a spatial system. Emerging rotating spirals composed by the competing species are frequently observed character of the resulting patterns. But how do these spiraling patterns change when we vary the external environment which affects the general vitality of individuals? Motivated by this question we suggest an off-lattice version of the tradition May--Leonard model which allows us to change the actual state of the environment gradually. This can be done by introducing a local carrying capacity parameter which value can be varied gently in an off-lattice environment. Our results support a previous analysis obtained in a more intricate metapopulation model and we show that the well-known rotating spirals become evident in a benign environment when the general density of the population is high. The accompanying time-dependent oscillation of competing species can also be detected where the amplitude and the frequency show a scaling law of the parameter that characterizes the state of the environment. These observations highlight that the assumed non-transitive interaction alone is insufficient condition to maintain biodiversity safely, but the actual state of the environment, which characterizes the general living conditions, also plays a decisive role on the  evolution of related systems.
\end{abstract}
\begin{document}

\flushbottom
\maketitle

\thispagestyle{empty}

\section*{Introduction}

According to the Darwinian selection hypothesis only the most viable competitor should survive as a result of a selection process. But we witness an amazing diversity of species in nature, which begs alternative explanations in ecology and in other complex competitive systems. The presence of a cyclic dominance among competitors is an elegant and very simple clue to resolve this contradiction. Indeed, scientists have observed several cases in living systems where the mentioned type of interaction can be observed. Examples can be given from microbial and plant communities, to coral reef, lizards, salmons and human interactions \cite{garde_rsob20,cameron_jecol09,jackson_pnas75,sinervo_n96,guill_jtb11,2014-Szolnoki-JRSI-11-0735}. We should stress, however, that similar cyclic dominance can also be detected in so-called social systems, where different strategies may dominate each other in a non-transitive way \cite{hauert_s02,szolnoki_csf20b}.

The basic model describing such kind of interactions between system members is based on the well-known rock-scissors-paper game where every member is a predator of another member and a prey for the third one simultaneously. Naturally, the strength of the dominance could be different between some predator-prey pairs and this asymmetry can provide some counter-intuitive system behaviors. One of them is the so-called ``survival of the weakest'' effect where the species having the lowest invasion rate develops the highest fraction in the population \cite{1991-Tainaka-EPL-15-399,2001-Frean-PRSLB-268-1323}. This phenomenon was confirmed in several modified models during the years \cite{2009-Berr-PRL-102-048102, szolnoki_csf20b,2020-Bazeia-CSF-141-110356, Liao2020} and in general the related dynamical behavior of cyclically dominant systems has collected significant research interest in the last decade \cite{baker_jtb20,nagatani_c20,2019-Brown-PRE-99-062116,2017-Lutz-Games-8-10,2018-Park-Chaos-28-113110,2016-Roman-JTB-403-10,palombi_epjb20,nagatani_srep18}. For example, the $n$ number of cyclically competing species can be extended to an arbitrary number where $n$ plays a central role on the resulting dynamics. This generalized version has a very rich structure leading to the formation of multi-domains of one or more species, which are separated by interfaces \cite{2012-Avelino-PRE-86-036112,roman_jsm12}. Also, the increase of the number of species usually leads to the development of more complex dynamical patterns. By focusing on the interplay between competition and partnership in spatial environments it can be observed that the development of neutral associations between individuals belonging to enemy partnership seems to affect the development of the dynamical structure along interfaces separating competing domains. For further details and a general overview of the present state of this research avenue we refer the interested reader to recent review papers \cite{2018-Dobramysl-JPA-51-063001,szolnoki_epl20,broom_dga21}.

Technically the related problems can be studied in Lotka--Volterra and May--Leonard models where the spatial distribution of species have a decisive role on the final evolutionary outcomes \cite{nagatani_pa19b,roman_pre13,he_q_epjb11,szolnoki_njp15}. In Lotka--Volterra models the application of $3\leq n$ species offers the simplest extension of a rock-scissors-papers-like cyclic dominance where predation and reproduction may occur in an elementary process \cite{frachebourg_pre96,szabo_pre08,park_c19b}. 
In a May--Leonard model the cyclic invasion is split into a ``selection" and a probabilistic reproduction step which makes the sum of all individuals a non-conserved quantity.
Strongly related to the scope of our present study, it turned out that mobility has a decisive role on the evolving pattern \cite{2007-Reichenbach-N-488-1046,reichenbach_jtb08,peltomaki_pre08}. More precisely, when the typical length of rotating spirals become comparable to the system size due to strong diffusion then the system can easily evolves into a trapped, or absorbing state where only a single species survives.

The above mentioned unequal invasion rates could be the result of an environmental factor, which is in general a parallel research avenue in complex systems. Heterogeneous environment can modify dynamical process directly \cite{chen_xj_pre09b,gracia-lazaro_csf13,wu_t_epl09,xia_cy_acs12,yang_lh_csf21,esmaeili_pre18,shao_yx_epl19}, which could be a local or a seasonal, or time-dependent change \cite{szolnoki_srep19,taitelbaum_prl20}. But we may control the state of the environment to modify the fractions of competing agents intentionally \cite{jansen_mb05,szolnoki_epl17,xie_yy_pa18}. Furthermore, the actual state of the environment can determine the vitality of the whole population fundamentally because adverse conditions may prevent individuals to survive while beneficial environment with unlimited resources can offer optimal living conditions, hence supports species. A natural question is how general environmental conditions influence the established cyclic dominance in the whole population. Is there any consequence on the evolving patterns when the environment makes easy or difficult for species to reproduce? An extreme case could be when a possible death of the individuals due to starvation is considered, which happens when a certain individual fails a given number of times when attempting predation \cite{2019-Avelino-EPL-126-68002}. In this case it was observed that the death of these individuals provide a crucial contribution to preservation of coexistence.

As a first step toward a more comprehensive understanding, in the following we study a model where the general state of the environment is modeled via a single
parameter which determines the local carrying capacity of the system. In this way we can vary the living conditions of all competitors uniformly and monitor how such changes influence the resulting evolutionary outcome. We note, however, that similar question was raised by other scientists previously who studied a well-mixed system or a spatially structured metapopulation \cite{west_jtb20,2014-Szczesny-PRE-90-032704}. For our present study it is important to stress that emerging rotating spirals and spiral waves are frequently observed accompanying patterns of cyclic dominance in spatial systems \cite{2013-Szczesny-EPL-102-28012,mobilia_g16,2014-Szczesny-PRE-90-032704,frey_pa10,szabo_pre99}. Therefore it is a fundamental question to study these arrangement when external conditions are varied. From this viewpoint there is a crucial technical circumstance that need to be mentioned. Generally, the application of lattice-type interaction topology makes simulations significantly easier, while the most important behaviors are still observable in these systems \cite{2004-Szabo-JPAMG-31-2599, 2009-Zhang-PRE-79-062901, 2014-Laird-Oikos-123-472, 2014-Rulquin-PRE-89-032133}. However, there is a drawback of the mentioned modeling technique in our present case which has a paramount importance. In particular, a lattice topology allows to change external conditions by discrete steps only. An alternative technique could be the so-called off-lattice simulations where the positions of individuals, hence their neighborhood may change continuously \cite{de-oliveira_csf21, 2020-Bazeia-EPL-129-28002,2010-Ni-C-20-045116,2010-Ni-PRE-82-066211}. The latter makes us possible to tune system parameters almost continuously, hence the control parameter which characterizes of the status of the environment can be varied finely. Indeed, the latter technique is more demanding and requires larger numerical efforts, but in certain cases we cannot avoid this difficulty. For example this is the proper way to study certain phenomena, like clustering \cite{2020-Bazeia-EPL-129-28002,de-oliveira_pa21}. 

As we will show, the emerging spatio-temporal pattern depends sensitively on the actual state of the external environment which directly determines the general living condition of the whole population.  In what follows, we first present the suggested off-lattice version of the May--Leonard model and its mathematical details. We then proceed with the presentation of the main results and followed by a discussion of their wider implications.

\section*{Model specification}

In this paper we shall consider a square box of linear size $L=1$ with periodic boundary conditions, which is the stage of our off-lattice simulations within the framework of a May--Leonard model. However, we should note that because of off-lattice character of the simulation the actual shape has no particular significance on the final outcomes. According to the traditional setup three different species, $A$, $B$ and $C$, are fighting for space where they dominate each others cyclically. In particular, the species $A$ preys the species $B$, that preys the species $C$ and the $C$ preys the species $A$, hence closing the cycle. The predation only occurs if there is a prey inside a circle of radius $\ell_p$ (predation length) centered around the predator and in this case the closest prey dies out. If there is no prey within the circle, then nothing happens.  

An alternative elementary process is a prey-independent reproduction of the focal individual. In this case a successor emerges within a radius $\ell_m$, but only if the total number of all individuals within the reproduction range ($\ell_r$) is smaller than $M$. The latter parameter, called local carrying capacity, characterizes the actual state of the environment. In a harsh environment the value of $M$ is low because limited resources cannot keep more individuals alive, while in a benign environment this value is higher. The third microscopic process is individual movement. In this case the chosen individual moves by a distance $\ell_m$ (movement length) in a randomly chosen direction. Notably, this step is always executed, while the success of predation and reproduction may depend on other circumstances, like the presence of a prey, or the total number of individuals in the reproduction neighborhood.

Initially $n_A = n_B = n_C = 10^4$ individuals of species are distributed randomly where the horizontal and vertical coordinates of every individuals are continuous variables. The total number of the whole population is $N = n_A + n_B + n_C$, which may change in time due to the above mentioned stochastic processes. It is worth noting that when we calculate the proper distance of two individuals we consider the mentioned periodic boundary conditions.

During an elementary Monte Carlo (MC) step an active individual is chosen randomly, which may move, predate or reproduce. The related probabilities are $m$, $p$ and $r$, respectively. For a full MC step we repeat the elementary steps $N$ times. In the following we have chosen $m=0.5$ and $p=r=0.25$ unless otherwise stated. Furthermore, for the characteristic lengths we used $\ell_p=\ell_r=0.02$, and $\ell_m= 0.01$. These values allow us to observe proper behavior of the spatial system. But we stress that similar observations can be made if we use other values of our model parameters therefore in the following we present characteristic and typical system reactions in dependence of environmental changes. We should also emphasize that by choosing too large length values, when $\ell \thickapprox 1$, the scales of microscopic steps become comparable to the system size. In this extreme case we would terminate onto a well-mixed system where the actual spatial distribution of individuals has no particular importance. To obtain the expected accuracy for the above mentioned parameter values we have repeated every run $10^3$ times by using independent initial conditions and averaged the individual results. Further details of our numerical experiments are given in the next section.

\section*{Results}

\begin{figure}[ht]
\centering
\includegraphics[scale=1.4]{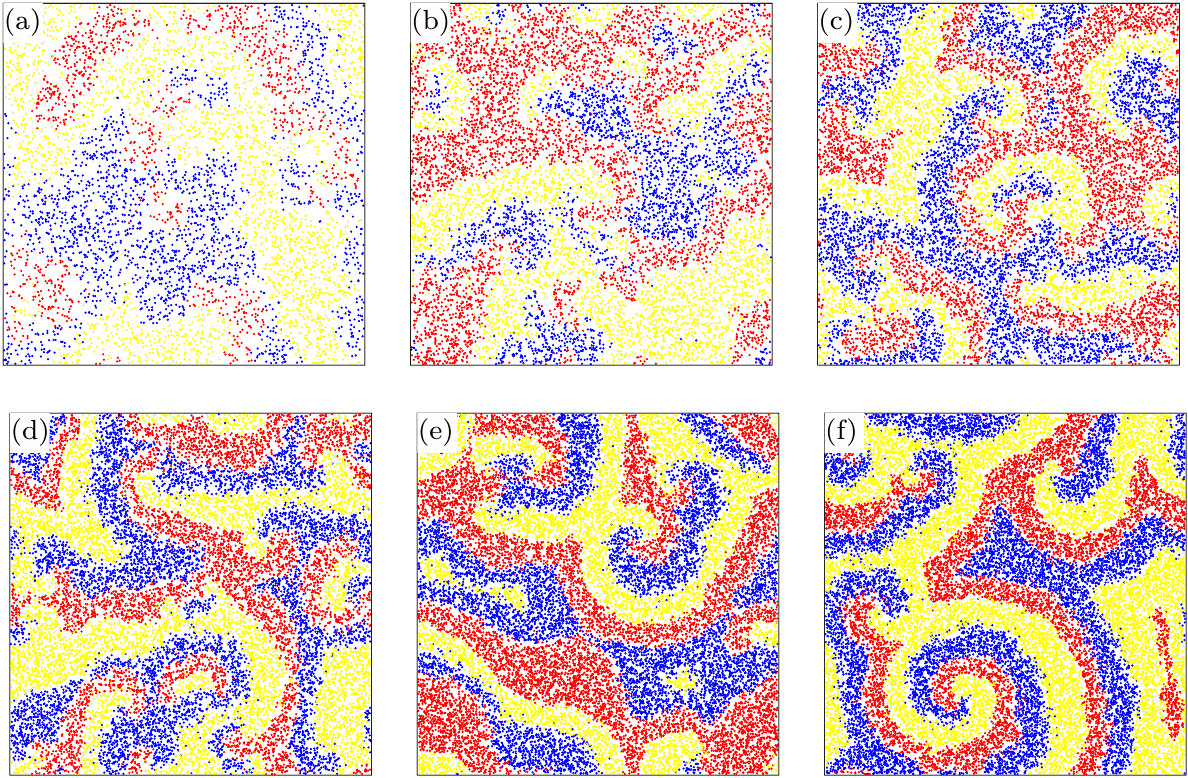}
\caption{Characteristic spatial distributions of species, for different values of the local carrying capacity $M$. Here red, blue, and yellow dots represent the three competing species, while the empty space is marked by white. The snapshots were taken after $10^8$ MC steps. The actual values of $M$ is 5 (a), 10 (b), 15 (c), 20 (d), 25 (e), and 30 (f), respectively. The comparison highlights clearly that the rotating spiral pattern becomes more evident as $M$ is increased.}
\label{snapshots}
\end{figure}

We first present a general overview about the impact of environmental change on the emerging spatial pattern of competing species. Figure~\ref{snapshots} depicts six representative snapshots of our off-lattice May--Leonard-model for different values of local carrying capacity $M$. The competing $A$, $B$, and $C$ species are marked by red, blue, and yellow colors respectively, while white marks empty space. The comparison illustrates very clearly that the rotating spirals become more evident as we increase the value of $M$. In parallel, the total number of individuals also increases by increasing $M$ because the portion of white color becomes gradually smaller.

\begin{figure}
\centering
\includegraphics[scale=1.1]{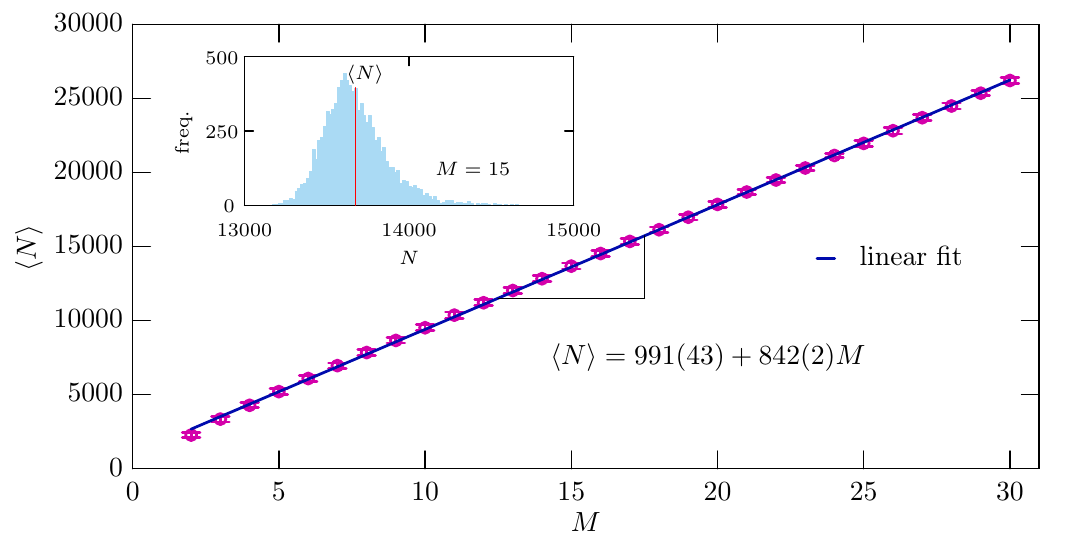}
\caption{The $\langle N \rangle$ average number of individuals in dependence of the local carrying capacity $M$. The main plot suggests that the average is proportional to $M$. The actual $N$ number of individuals, however, fluctuates heavily in time. The inset depicts a typical distribution of $N$ values collected from $4 \cdot 10^4$ MC steps after $10^8$ MC relaxation steps when $M=15$. The calculated average $\langle N \rangle$ is marked by a red line in the inset.}
\label{total}
\end{figure}

Our last observation is summarized in Fig.~\ref{total} in a more quantitative way. Here we plot the $\langle N \rangle$ average number of individuals in the whole population in dependence of $M$. We stress that this is an average value because the temporary number of individuals may change in time. To illustrate it, in the inset we show a representative distribution of $N$ in the stationary state for $M=15$. Here the position of the calculated average is marked by a vertical red line. Note that the error bars are also marked in the main plot and to obtain the requested accuracy we calculated the time average over $4 \cdot 10^4$ MC steps after $10^8$ MC steps of relaxation for each $M$ value. We note that the actual number of iterations depends on the size of population, therefore to obtain the related data for higher $M$ requires significantly larger numerical efforts. As the main plot suggests, the average value follows a linear dependence on $M$ where the value of the slope is quite robust and does not depend on microscopic details, like the $\ell_m$ value. In particular, if we allow more intensive individual movement, for instance, then the increment of total individuals will change in a similar way, having the same slope, as we increase $M$. 

As we already noted, we have also explored what happens in the extreme cases when a characteristic length of movement becomes comparable to the system size. For example when $\ell_m$ is increased too long, say $\ell_m =0.8$ then the typical domain sizes grow which shows similar behavior can be observed in a well-mixed system. In this case the fluctuations can be so large that one of the competing species goes extinct, which breaks the symmetry and the population eventually terminates to a homogeneous state. This behavior is similar to those previously observed on a lattice structure \cite{2007-Reichenbach-N-488-1046}. Another interesting behavior can be observed when the predation and reproduction lengths become comparable to the system size. In this case empty space behaves like an ``additional'' species and the real species become seemingly isolated from each other. As a result, the deserted areas occupy a significant portion of available space. We will discuss this phenomenon later but first we focus on other aspects of emerging spirals.

\begin{figure}[ht]
\centering
\includegraphics[scale=1.1]{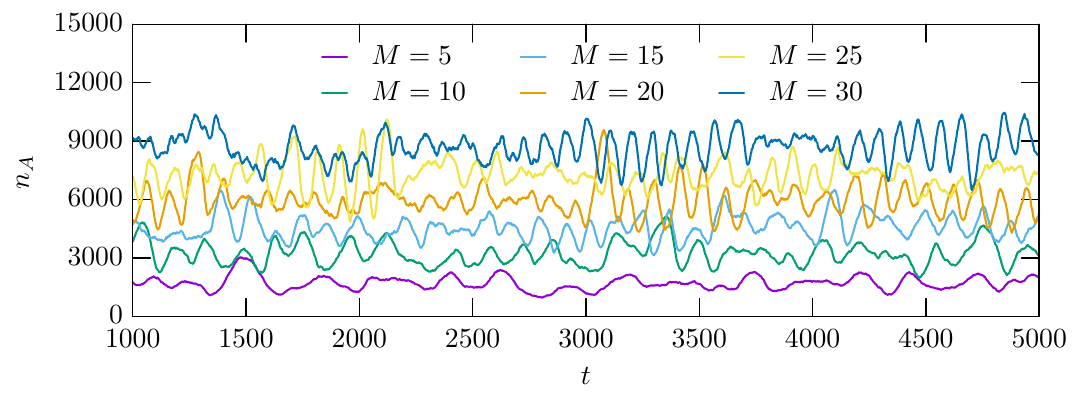}
\caption{Time dependence of the number of individuals of species $A$ for different values of the local carrying capacity. The applied $M$ values are shown in the legend. These lines suggest that the oscillation becomes pronounced for higher $M$ values. This behavior agrees with the one we observed in Fig.~\ref{snapshots}, indicating the emergency of rotating spirals and waves. Naturally, similar patterns can be detected for the remaining two species because of the symmetry of the model.}
\label{osc}

\end{figure}

The emergence of rotating spirals has a detectable consequence not only on spatial, but also on temporal patterns. The latter fact can be illustrated properly if we monitor the time dependence of the fraction of a certain species. This phenomenon is shown in Fig.~\ref{osc} where we plot the temporary number of individuals for species $A$ for different values of $M$. As the plot suggests the oscillation becomes more intensive for larger $M$ values. Obviously, similar patterns can be obtained for the remaining two species because the non-transitive interaction establishes a symmetry among the competing species. While for small $M$ values, when the environment is harsh, the time course seems to be noisy, but for high values of $M$ the environment becomes rich of resources hence it is capable to maintain a large populations stably. This can also be read out from the plot because the average level of $n_A$ increases gradually as we increase the value of $M$. There is, however, an important feature of the time dependence which underlines the main conclusion of our study. In a stochastic simulation it is a generally expected behavior that for larger population the system behavior becomes less noisy. Indeed, this also happens in our model and the oscillation becomes more regular as we reach higher $M$ values, hence indirectly enlarge the size of the whole population. The amplitude of the oscillation, however remains significant hence indicating the emergence of spirals and waves we already have shown in Fig.~\ref{snapshots}. 

\begin{figure}[ht]
\centering
\includegraphics[scale=1.1]{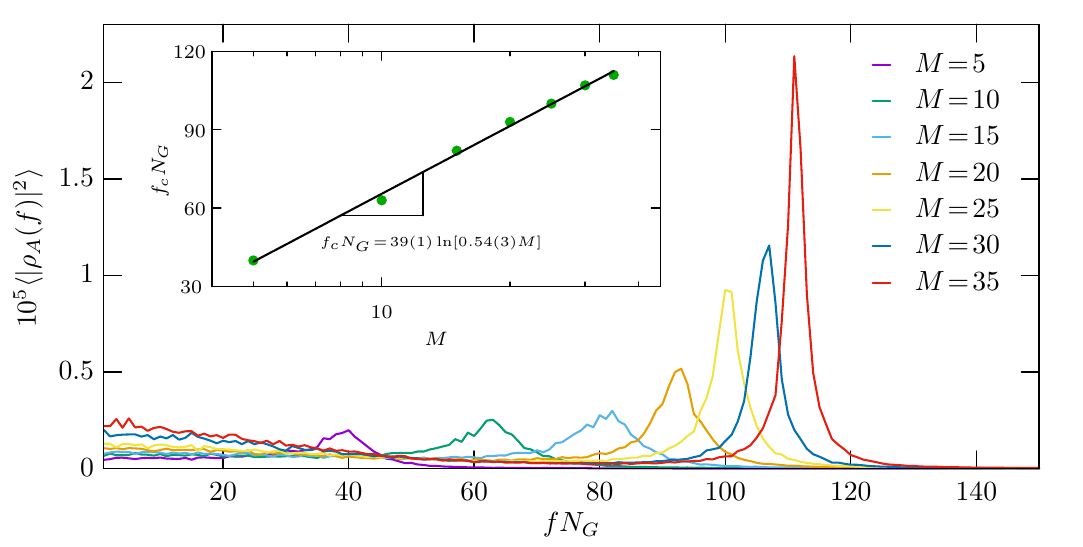}
\caption{The power spectrum calculated from time dependence of the fraction of species $A$ as a function of frequency that is multiplied by the number of steps. The curves are related to different values of local carrying capacity $M$, as indicated by the legend. All curves has a peak which position determines a characteristic frequency. The insert shows a clear logarithmic peak of this characteristic frequency as a function of $M$. Similar plots can be obtained for the remaining two species. This plot confirms quantitatively that the time-dependent oscillation as an accompanying feature of rotating spirals become conspicuous for high $M$ when the environment is capable to maintain a rich population.}
\label{peak}
\end{figure}

To give a more quantitative description about the oscillations we apply the Fourier analysis to the $\rho(t)$ function which determines the fraction of a species in time. The temporal discrete Fourier transform can be given as
\begin{equation}
	\rho(f) = \dfrac{1}{N_G} \displaystyle \sum_{t=0}^{N_G-1} \rho(t)\cdot e^{-2\pi ift} \ ,
	\label{eq1}
\end{equation}
where the coefficient of $f$ is calculated from $N_G=10^4$ components. To collect reliable data for the stationary states we always used 1000 MC steps of relaxation. The resulting power spectrum $\langle |\rho_A(f)|^2\rangle$ of species $A$ is shown in Fig.~\ref{peak} where we plotted the curves obtained for different $M$ values simultaneously. These values are indicated in the legend. We note that for an appropriate scale all power spectrum values are multiplied by a $10^5$ constant factor. For the requested accuracy we averaged the data over 250 independent simulations where the system evolution was launched from different initial states. Evidently, similar curves can be obtained for the remaining $B$ and $C$ species. The comparison of curves indicates that the location of the peak shifts toward higher frequency values as we increase $M$. For example for $M = 30$ we have 107 oscillations during $10^4$ Monte Carlo steps. In parallel, the height of the peak also increases by enlarging $M$, which suggests that the oscillation becomes more characteristic as the living conditions are improved. Notably, the position of the peaks show a nice logarithmic scaling as it is shown in the inset of Fig.~\ref{peak}. This quantitative analysis confirms what we already observed in Fig.~\ref{osc}. Namely, even if there is a clear non-transitive cyclic interaction among competing species, the well-known rotating spirals and accompanying oscillations are hardly detectable if the environment is poor of resources and can only maintain a stunted population. We stress that the biodiversity is still maintained, but not in the presence of rotating spirals we frequently expect from a spatial system having cyclic dominant microscopic dynamic. However, if the living conditions are improved then the anticipated rotating spirals of spatial distribution and the time-dependent oscillation of species become evident again.    

\begin{figure}[ht]
\centering
\includegraphics[scale=1.4]{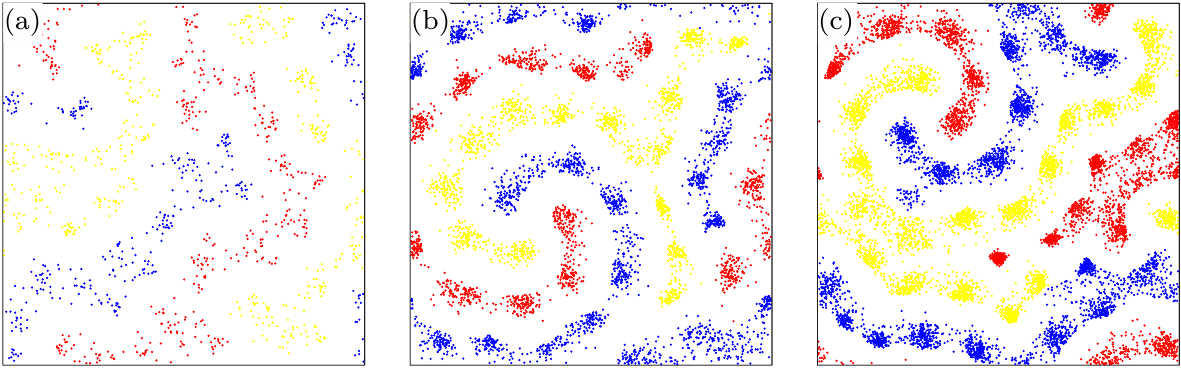}
\caption{Characteristic spatial distribution of species at $\ell_p=\ell_r=0.1$, $\ell_m=0.01$ parameter values. The only difference is we use $M=30$ on panel~(a), $M=120$ on panel~(b) and $M=240$ on panel~(c). This figure illustrates that alone the value of $M$ is insufficient to characterize the state of environment, because its combination with the length scales and the probabilities determining dynamical processes could tell us the proper living conditions. The increase of $M$, however, when all other parameters are fixed, can restore the rotating spirals we expect from a cyclically dominant spatial system. Interestingly, the large predation and reproduction ranges result in large portion of empty space which behaves as an additional inseparable part of the spirals.}
\label{large_rep}
\end{figure}

Before concluding we should highlight that alone the introduced carrying capacity parameter is just an initial step to model the general conditions of the environment. More precisely, the single value of $M$ does not determine the state of the environment accurately, because this parameter is linked to the reproduction length $\ell_r$. When the latter is large then even a relatively high $M$ value still represents a poor environment. This is demonstrated in Fig.~\ref{large_rep} where we applied five times larger reproduction and predation lengths as previously. Therefore, even if we used quite large $M$ values, the total sum of individuals remain low in the population. As an accompanying effect, the rotating spirals diminish from the pattern. But they can be recovered again if we increase $M$, as it is done in panels~(b) and (c) of Fig.~\ref{large_rep}. Interestingly, the portion of empty space remains high, and seems to behave as an organic additional member of the spirals. This is a phenomenon that cannot be observed in a cyclic system where the size of the population is strictly constant in time \cite{2014-Szczesny-PRE-90-032704}.

\section*{Discussion}

It is a well-known fact that cyclic dominance among competing species not only maintains diversity but frequently generates spiral waves in a spatial system where agents have limited access to other interacting partners. There are, however, some circumstances when this general picture is broken. An example could be when the high mobility of species destroys the above described patterns and jeopardizes the coexistence of all competitors \cite{2007-Reichenbach-N-488-1046}. Similarly, the breaking of unidirectional invasions, or heterogeneity in site-specific invasion rates could also terminate the coexistence of competing species \cite{2020-Bazeia-CSF-141-110356,szolnoki_srep16b}. External factors, like the proper state of environment, which generally determines the viability of the population, also seems to be a crucial ingredient to this problem. Motivated by this argument, in this work we explored how the actual state of environment influences directly the competition of equally strong opponents. In our simple model the mentioned condition can be controlled via a single parameter that determines the local carrying capacity of the system. The other central feature of our model was the off-lattice topology which made possible to vary this parameter and the related external condition gently.

Our first observation is the stable coexistence of competing species for low value of $M$ when the general maintaining capacity of the environment is moderate. But in this limit, spiral patterns, which characterizes the traditional three-species model, cannot be detected. In parallel, the time dependence of a certain species shows an irregular sequence. This anomaly, however, can be restored if we enlarge $M$, which involves a more supporting environment hence can maintain a more crowded population. It is worth noting that conceptually similar observation was analyzed by Szczesny, Mobilia, and Rucklidge, who considered a metapopulation model where every patches, which are organized in a square grid, have a limited carrying capacity \cite{2014-Szczesny-PRE-90-032704}. In their cases when the sizes of the well-mixed sub-populations were small the rotating spirals disappeared, but they were restored again by increasing the size of the mentioned sub-populations. The agreement between their partial differential equation approach and our present off-lattice simulations underlines the universality and broadens the validity of the presented observation.

We also demonstrated that there is a linear dependence of the average size of the population on the $M$ parameter and this slope is largely independent on microscopic details of the used May--Leonard model. We have also studied how a regular oscillation of time dependent fraction of species emerges as we improve the general quality of the environment. Furthermore, we demonstrated that the characteristic frequency of the oscillation shows a logarithmic scaling with the introduced $M$ parameter.

Summing up, our work pointed out that the quality of the environment, which provides a stage for competition of species and determines the general living conditions, can be a decisive factor for what kind of spatial and temporal patterns emerge. When the general living conditions are poor, because of the lack of resources or for other reasons, then the well-known rotating spirals characterizing such kind of spatial system are missing, but they become detectable again if the living conditions are improved. Similar behavior was already observed in a case when intensive diffusion resulted in the disappearance of spirals. This last effect, however, can be easily explained because fast individual movement helps more intensive mixing hence it drifts the spatial system toward the well-mixed behavior. But in our case the poor environment has not similar consequence therefore the lack of rotating spirals seems to be surprising at first sight. On the other hand the reported behavior is robust because other modeling approach, by assuming metapopulation setup, also confirmed it. Nevertheless, this phenomenon can also explain why we can observe spiral patterns rarely in field studies where cyclic dominance is present otherwise.

\section*{Acknowledgements}

B.F.O. thank Fundação Araucária, and INCT-FCx (CNPq/FAPESP) for financial and computational support. D.B. acknowledges Conselho Nacional de Desenvolvimento Cient\'\i fico e Tecnol\'ogico (CNPq, Grants nos. 303469/2019-6 and 404913/2018-0) and Para\'\i ba State Research Foundation (FAPESQ-PB, Grant no. 0015/2019) for financial support.

\section*{Author contributions statement}

D.B., B.F.O. and A.S. devised the research project. B.F.O. performed numerical simulations. M.J.B.F. and B.F.O. wrote the original draft. B.F.O. and A.S. analyzed the results. A.S. wrote the final version. All authors reviewed the manuscript.

\section*{Additional information}

\textbf{Competing interests} The authors declare no competing interests.


\begin{thebibliography}{10}
\urlstyle{rm}
\expandafter\ifx\csname url\endcsname\relax
  \def\url#1{\texttt{#1}}\fi
\expandafter\ifx\csname urlprefix\endcsname\relax\def\urlprefix{URL }\fi
\expandafter\ifx\csname doiprefix\endcsname\relax\def\doiprefix{DOI: }\fi
\providecommand{\bibinfo}[2]{#2}
\providecommand{\eprint}[2][]{\url{#2}}

\bibitem{garde_rsob20}
\bibinfo{author}{Garde, R.}, \bibinfo{author}{Ewald, J.},
  \bibinfo{author}{Kov{\'a}cs, {\'A}.~T.} \& \bibinfo{author}{Schuster, S.}
\newblock \bibinfo{journal}{\bibinfo{title}{Modelling population dynamics in a
  unicellular social organism community using a minimal model and evolutionary
  game theory}}.
\newblock {\emph{\JournalTitle{Open Biol.}}} \textbf{\bibinfo{volume}{10}},
  \bibinfo{pages}{200206} (\bibinfo{year}{2020}).

\bibitem{cameron_jecol09}
\bibinfo{author}{Cameron, D.~D.}, \bibinfo{author}{White, A.} \&
  \bibinfo{author}{Antonovics, J.}
\newblock \bibinfo{journal}{\bibinfo{title}{Parasite-grass-forb interactions
  and rock-paper-scissor dynamics: predicting the effects of the parasitic
  plant \protect{Rhinanthus} minor on host plant communities}}.
\newblock {\emph{\JournalTitle{J. Ecol.}}} \textbf{\bibinfo{volume}{97}},
  \bibinfo{pages}{1311--1319} (\bibinfo{year}{2009}).

\bibitem{jackson_pnas75}
\bibinfo{author}{Jackson, J. B.~C.} \& \bibinfo{author}{Buss, L.}
\newblock \bibinfo{journal}{\bibinfo{title}{Allelopathy and spatial competition
  among coral reef invertebrates}}.
\newblock {\emph{\JournalTitle{Proc. Nat. Acad. Sci. USA}}}
  \textbf{\bibinfo{volume}{72}}, \bibinfo{pages}{5160--5163}
  (\bibinfo{year}{1975}).

\bibitem{sinervo_n96}
\bibinfo{author}{Sinervo, B.} \& \bibinfo{author}{Lively, C.~M.}
\newblock \bibinfo{journal}{\bibinfo{title}{The rock-paper-scissors game and
  the evolution of alternative male strategies}}.
\newblock {\emph{\JournalTitle{Nature}}} \textbf{\bibinfo{volume}{380}},
  \bibinfo{pages}{240--243} (\bibinfo{year}{1996}).

\bibitem{guill_jtb11}
\bibinfo{author}{Guill, C.}, \bibinfo{author}{Drossel, B.},
  \bibinfo{author}{Just, W.} \& \bibinfo{author}{Carmack, E.}
\newblock \bibinfo{journal}{\bibinfo{title}{A three-species model explaining
  cyclic dominance of \protect{Pacific} salmon}}.
\newblock {\emph{\JournalTitle{J. Theor. Biol.}}}
  \textbf{\bibinfo{volume}{276}}, \bibinfo{pages}{16--21}
  (\bibinfo{year}{2011}).

\bibitem{2014-Szolnoki-JRSI-11-0735}
\bibinfo{author}{Szolnoki, A.} \emph{et~al.}
\newblock \bibinfo{journal}{\bibinfo{title}{Cyclic dominance in evolutionary
  games: a review}}.
\newblock {\emph{\JournalTitle{J. R. Soc. Interface}}}
  \textbf{\bibinfo{volume}{11}}, \bibinfo{pages}{20140735}
  (\bibinfo{year}{2014}).

\bibitem{hauert_s02}
\bibinfo{author}{Hauert, C.}, \bibinfo{author}{De~Monte, S.},
  \bibinfo{author}{Hofbauer, J.} \& \bibinfo{author}{Sigmund, K.}
\newblock \bibinfo{journal}{\bibinfo{title}{Volunteering as \protect{Red Queen}
  mechanism for cooperation in public goods game}}.
\newblock {\emph{\JournalTitle{Science}}} \textbf{\bibinfo{volume}{296}},
  \bibinfo{pages}{1129--1132} (\bibinfo{year}{2002}).

\bibitem{szolnoki_csf20b}
\bibinfo{author}{Szolnoki, A.} \& \bibinfo{author}{Chen, X.}
\newblock \bibinfo{journal}{\bibinfo{title}{Strategy dependent learning
  activity in cyclic dominant systems}}.
\newblock {\emph{\JournalTitle{Chaos, Solitons \& Fractals}}}
  \textbf{\bibinfo{volume}{138}}, \bibinfo{pages}{109935}
  (\bibinfo{year}{2020}).

\bibitem{1991-Tainaka-EPL-15-399}
\bibinfo{author}{Tainaka, K.} \& \bibinfo{author}{Itoh, Y.}
\newblock \bibinfo{journal}{\bibinfo{title}{Topological phase transition in
  biological ecosystems}}.
\newblock {\emph{\JournalTitle{Europhysics Letters}}}
  \textbf{\bibinfo{volume}{15}}, \bibinfo{pages}{399--404}
  (\bibinfo{year}{1991}).

\bibitem{2001-Frean-PRSLB-268-1323}
\bibinfo{author}{Frean, M.} \& \bibinfo{author}{Abraham, E.~R.}
\newblock \bibinfo{journal}{\bibinfo{title}{Rock-scissors-paper and the
  survival of the weakest}}.
\newblock {\emph{\JournalTitle{Proc. R. Soc. Lond. B}}}
  \textbf{\bibinfo{volume}{268}}, \bibinfo{pages}{1323--1327}
  (\bibinfo{year}{2001}).

\bibitem{2009-Berr-PRL-102-048102}
\bibinfo{author}{Berr, M.}, \bibinfo{author}{Reichenbach, T.},
  \bibinfo{author}{Schottenloher, M.} \& \bibinfo{author}{Frey, E.}
\newblock \bibinfo{journal}{\bibinfo{title}{Zero-one survival behavior of
  cyclically competing species}}.
\newblock {\emph{\JournalTitle{Phys. Rev. Lett.}}}
  \textbf{\bibinfo{volume}{102}}, \bibinfo{pages}{048102}
  (\bibinfo{year}{2009}).

\bibitem{2020-Bazeia-CSF-141-110356}
\bibinfo{author}{Bazeia, D.}, \bibinfo{author}{de~Oliveira, B.},
  \bibinfo{author}{Silva, J.} \& \bibinfo{author}{Szolnoki, A.}
\newblock \bibinfo{journal}{\bibinfo{title}{Breaking unidirectional invasions
  jeopardizes biodiversity in spatial \protect{May--Leonard} systems}}.
\newblock {\emph{\JournalTitle{Chaos, Solitons \& Fractals}}}
  \textbf{\bibinfo{volume}{141}}, \bibinfo{pages}{110356}
  (\bibinfo{year}{2020}).

\bibitem{Liao2020}
\bibinfo{author}{Liao, M.~J.}, \bibinfo{author}{Miano, A.},
  \bibinfo{author}{Nguyen, C.~B.}, \bibinfo{author}{Chao, L.} \&
  \bibinfo{author}{Hasty, J.}
\newblock \bibinfo{journal}{\bibinfo{title}{Survival of the weakest in
  non-transitive asymmetric interactions among strains of e. coli}}.
\newblock {\emph{\JournalTitle{Nat. Commun.}}} \textbf{\bibinfo{volume}{11}},
  \bibinfo{pages}{6055} (\bibinfo{year}{2020}).

\bibitem{baker_jtb20}
\bibinfo{author}{Baker, R.} \& \bibinfo{author}{Pleimling, M.}
\newblock \bibinfo{journal}{\bibinfo{title}{The effect of habitats and fitness
  on species coexistence in systems with cyclic dominance}}.
\newblock {\emph{\JournalTitle{J. Theor. Biol.}}}
  \textbf{\bibinfo{volume}{486}}, \bibinfo{pages}{110084}
  (\bibinfo{year}{2020}).

\bibitem{nagatani_c20}
\bibinfo{author}{Nagatani, T.} \& \bibinfo{author}{Ichinose, G.}
\newblock \bibinfo{journal}{\bibinfo{title}{Diffusively-coupled
  rock-paper-scissors game with mutation in scale-free hierarchical networks}}.
\newblock {\emph{\JournalTitle{Complexity}}} \textbf{\bibinfo{volume}{2020}},
  \bibinfo{pages}{6976328} (\bibinfo{year}{2020}).

\bibitem{2019-Brown-PRE-99-062116}
\bibinfo{author}{Brown, B.~L.}, \bibinfo{author}{Meyer-Ortmanns, H.} \&
  \bibinfo{author}{Pleimling, M.}
\newblock \bibinfo{journal}{\bibinfo{title}{Dynamically generated hierarchies
  in games of competition}}.
\newblock {\emph{\JournalTitle{Phys. Rev. E}}} \textbf{\bibinfo{volume}{99}},
  \bibinfo{pages}{062116} (\bibinfo{year}{2019}).

\bibitem{2017-Lutz-Games-8-10}
\bibinfo{author}{L\"{u}tz, A.}, \bibinfo{author}{Cazaubiel, A.} \&
  \bibinfo{author}{Arenzon, J.}
\newblock \bibinfo{journal}{\bibinfo{title}{Cyclic competition and percolation
  in grouping predator-prey populations}}.
\newblock {\emph{\JournalTitle{Games}}} \textbf{\bibinfo{volume}{8}},
  \bibinfo{pages}{10} (\bibinfo{year}{2017}).

\bibitem{2018-Park-Chaos-28-113110}
\bibinfo{author}{Park, J.}, \bibinfo{author}{Do, Y.} \& \bibinfo{author}{Jang,
  B.}
\newblock \bibinfo{journal}{\bibinfo{title}{Multistability in the cyclic
  competition system}}.
\newblock {\emph{\JournalTitle{Chaos}}} \textbf{\bibinfo{volume}{28}},
  \bibinfo{pages}{113110} (\bibinfo{year}{2018}).

\bibitem{2016-Roman-JTB-403-10}
\bibinfo{author}{Roman, A.}, \bibinfo{author}{Dasgupta, D.} \&
  \bibinfo{author}{Pleimling, M.}
\newblock \bibinfo{journal}{\bibinfo{title}{A theoretical approach to
  understand spatial organization in complex ecologies}}.
\newblock {\emph{\JournalTitle{Journal of Theoretical Biology}}}
  \textbf{\bibinfo{volume}{403}}, \bibinfo{pages}{10--16}
  (\bibinfo{year}{2016}).

\bibitem{palombi_epjb20}
\bibinfo{author}{Palombi, F.}, \bibinfo{author}{Ferriani, S.} \&
  \bibinfo{author}{Toti, S.}
\newblock \bibinfo{journal}{\bibinfo{title}{Coevolutionary dynamics of a
  variant of the cyclic \protect{Lotka--Volterra} model with three-agent
  interactions}}.
\newblock {\emph{\JournalTitle{Eur. Phys. J. B}}}
  \textbf{\bibinfo{volume}{93}}, \bibinfo{pages}{194} (\bibinfo{year}{2020}).

\bibitem{nagatani_srep18}
\bibinfo{author}{Nagatani, T.}, \bibinfo{author}{Ichinose, G.} \&
  \bibinfo{author}{i.~Tainaka, K.}
\newblock \bibinfo{journal}{\bibinfo{title}{Heterogeneous network promotes
  species coexistence: metapopulation model for rock-paper-scissors game}}.
\newblock {\emph{\JournalTitle{Sci. Rep.}}} \textbf{\bibinfo{volume}{8}},
  \bibinfo{pages}{7094} (\bibinfo{year}{2018}).

\bibitem{2012-Avelino-PRE-86-036112}
\bibinfo{author}{Avelino, P.~P.}, \bibinfo{author}{Bazeia, D.},
  \bibinfo{author}{Losano, L.}, \bibinfo{author}{Menezes, J.} \&
  \bibinfo{author}{de~Oliveira, B.~F.}
\newblock \bibinfo{journal}{\bibinfo{title}{Junctions and spiral patterns in
  generalized rock-paper-scissors models}}.
\newblock {\emph{\JournalTitle{Phys. Rev. E}}} \textbf{\bibinfo{volume}{86}},
  \bibinfo{pages}{036112} (\bibinfo{year}{2012}).

\bibitem{roman_jsm12}
\bibinfo{author}{Roman, A.}, \bibinfo{author}{Konrad, D.} \&
  \bibinfo{author}{Pleimling, M.}
\newblock \bibinfo{journal}{\bibinfo{title}{Cyclic competition of four species:
  domains and interfaces}}.
\newblock {\emph{\JournalTitle{J. Stat. Mech.}}}
  \textbf{\bibinfo{volume}{2012}}, \bibinfo{pages}{P07014}
  (\bibinfo{year}{2012}).

\bibitem{2018-Dobramysl-JPA-51-063001}
\bibinfo{author}{Dobramysl, U.}, \bibinfo{author}{Mobilia, M.},
  \bibinfo{author}{Pleimling, M.} \& \bibinfo{author}{Täuber, U.~C.}
\newblock \bibinfo{journal}{\bibinfo{title}{Stochastic population dynamics in
  spatially extended predator–prey systems}}.
\newblock {\emph{\JournalTitle{Journal of Physics A: Mathematical and
  Theoretical}}} \textbf{\bibinfo{volume}{51}}, \bibinfo{pages}{063001}
  (\bibinfo{year}{2018}).

\bibitem{szolnoki_epl20}
\bibinfo{author}{Szolnoki, A.}, \bibinfo{author}{\protect{de Oliveira}, B.~F.}
  \& \bibinfo{author}{Bazeia, D.}
\newblock \bibinfo{journal}{\bibinfo{title}{Pattern formations driven by cyclic
  interactions: A brief review of recent developments}}.
\newblock {\emph{\JournalTitle{EPL}}} \textbf{\bibinfo{volume}{131}},
  \bibinfo{pages}{68001} (\bibinfo{year}{2020}).

\bibitem{broom_dga21}
\bibinfo{author}{Broom, M.}, \bibinfo{author}{Erovenko, I.~V.}
  \& \bibinfo{author}{Rycht{\'a}{\v r}, J.}
\newblock \bibinfo{journal}{\bibinfo{title}{Modelling Evolution in Structured Populations Involving Multiplayer Interactions}}.
\newblock {\emph{\JournalTitle{Dyn. Games Appl.}}} \textbf{\bibinfo{volume}{11}},
  \bibinfo{pages}{270--293} (\bibinfo{year}{2021}).

\bibitem{nagatani_pa19b}
\bibinfo{author}{Nagatani, T.}
\newblock \bibinfo{journal}{\bibinfo{title}{Diffusively coupled
  \protect{Lotka--Volterra} system stabilized by heterogeneous graphs}}.
\newblock {\emph{\JournalTitle{Physica A}}} \textbf{\bibinfo{volume}{525}},
  \bibinfo{pages}{1114--1123} (\bibinfo{year}{2019}).

\bibitem{roman_pre13}
\bibinfo{author}{Roman, A.}, \bibinfo{author}{Dasgupta, D.} \&
  \bibinfo{author}{Pleimling, M.}
\newblock \bibinfo{journal}{\bibinfo{title}{Interplay between partnership
  formation and competition in generalized \protect{May--Leonard} game}}.
\newblock {\emph{\JournalTitle{Phys. Rev. E}}} \textbf{\bibinfo{volume}{87}},
  \bibinfo{pages}{032148} (\bibinfo{year}{2013}).

\bibitem{he_q_epjb11}
\bibinfo{author}{He, Q.}, \bibinfo{author}{Mobilia, M.} \&
  \bibinfo{author}{T{\"a}uber, U.~C.}
\newblock \bibinfo{journal}{\bibinfo{title}{Co-existence in the two-dimensional
  \protect{May--Leonard} model with random rates}}.
\newblock {\emph{\JournalTitle{Eur. Phys. J. B}}}
  \textbf{\bibinfo{volume}{82}}, \bibinfo{pages}{97--105}
  (\bibinfo{year}{2011}).

\bibitem{szolnoki_njp15}
\bibinfo{author}{Szolnoki, A.} \& \bibinfo{author}{Perc, M.}
\newblock \bibinfo{journal}{\bibinfo{title}{Vortices determine the dynamics of
  biodiversity in cyclical interactions with protection spillovers}}.
\newblock {\emph{\JournalTitle{New J. Phys.}}} \textbf{\bibinfo{volume}{17}},
  \bibinfo{pages}{113033} (\bibinfo{year}{2015}).

\bibitem{frachebourg_pre96}
\bibinfo{author}{Frachebourg, L.}, \bibinfo{author}{Krapivsky, P.~L.} \&
  \bibinfo{author}{Ben-Naim, E.}
\newblock \bibinfo{journal}{\bibinfo{title}{Spatial organization in cyclic
  \protect{Lotka--Volterra} systems}}.
\newblock {\emph{\JournalTitle{Phys. Rev. E}}} \textbf{\bibinfo{volume}{54}},
  \bibinfo{pages}{6186--6200} (\bibinfo{year}{1996}).

\bibitem{szabo_pre08}
\bibinfo{author}{Szab{\'o}, G.} \& \bibinfo{author}{Szolnoki, A.}
\newblock \bibinfo{journal}{\bibinfo{title}{Phase transitions induced by
  variation of invasion rates in spatial cyclic predator-prey models with four
  or six species}}.
\newblock {\emph{\JournalTitle{Phys. Rev. E}}} \textbf{\bibinfo{volume}{77}},
  \bibinfo{pages}{011906} (\bibinfo{year}{2008}).

\bibitem{park_c19b}
\bibinfo{author}{Park, J.} \& \bibinfo{author}{Jang, B.}
\newblock \bibinfo{journal}{\bibinfo{title}{Robust coexistence with alternative
  competition strategy in the spatial cyclic game of five species}}.
\newblock {\emph{\JournalTitle{Chaos}}} \textbf{\bibinfo{volume}{29}},
  \bibinfo{pages}{051105} (\bibinfo{year}{2019}).

\bibitem{2007-Reichenbach-N-488-1046}
\bibinfo{author}{Reichenbach, T.}, \bibinfo{author}{Mobilia, M.} \&
  \bibinfo{author}{Frey, E.}
\newblock \bibinfo{journal}{\bibinfo{title}{Mobility promotes and jeopardizes
  biodiversity in rock-paper-scissors games}}.
\newblock {\emph{\JournalTitle{Nature}}} \textbf{\bibinfo{volume}{448}},
  \bibinfo{pages}{1046--1049} (\bibinfo{year}{2007}).

\bibitem{reichenbach_jtb08}
\bibinfo{author}{Reichenbach, T.}, \bibinfo{author}{Mobilia, M.} \&
  \bibinfo{author}{Frey, E.}
\newblock \bibinfo{journal}{\bibinfo{title}{Self-organization of mobile
  populations in cyclic competititon}}.
\newblock {\emph{\JournalTitle{J. Theor. Biol.}}}
  \textbf{\bibinfo{volume}{254}}, \bibinfo{pages}{368--383}
  (\bibinfo{year}{2008}).

\bibitem{peltomaki_pre08}
\bibinfo{author}{Peltom{\"a}ki, M.} \& \bibinfo{author}{Alava, M.}
\newblock \bibinfo{journal}{\bibinfo{title}{Three- and four-state
  rock-paper-scissors games with diffusion}}.
\newblock {\emph{\JournalTitle{Phys. Rev. E}}} \textbf{\bibinfo{volume}{78}},
  \bibinfo{pages}{031906} (\bibinfo{year}{2008}).

\bibitem{chen_xj_pre09b}
\bibinfo{author}{Chen, X.}, \bibinfo{author}{Fu, F.} \& \bibinfo{author}{Wang,
  L.}
\newblock \bibinfo{journal}{\bibinfo{title}{Social tolerance allows cooperation
  to prevail in an adaptive environment}}.
\newblock {\emph{\JournalTitle{Phys. Rev. E}}} \textbf{\bibinfo{volume}{80}},
  \bibinfo{pages}{051104} (\bibinfo{year}{2009}).

\bibitem{gracia-lazaro_csf13}
\bibinfo{author}{Gracia-L{\'a}zaro, C.}, \bibinfo{author}{Flor{\' \i}a, L.~M.},
  \bibinfo{author}{G{\'o}mez-Garde{\~n}es, J.} \& \bibinfo{author}{Moreno, Y.}
\newblock \bibinfo{journal}{\bibinfo{title}{Cooperation in changing
  environments: Irreversibility in the transition to cooperation in complex
  networks}}.
\newblock {\emph{\JournalTitle{Chaos, Solitons \& Fractals}}}
  \textbf{\bibinfo{volume}{56}}, \bibinfo{pages}{188--193}
  (\bibinfo{year}{2013}).

\bibitem{wu_t_epl09}
\bibinfo{author}{Wu, T.}, \bibinfo{author}{Fu, F.} \& \bibinfo{author}{Wang,
  L.}
\newblock \bibinfo{journal}{\bibinfo{title}{Individual's expulsion to nasty
  environment promotes cooperation in public goods games}}.
\newblock {\emph{\JournalTitle{EPL}}} \textbf{\bibinfo{volume}{88}},
  \bibinfo{pages}{30011} (\bibinfo{year}{2009}).

\bibitem{xia_cy_acs12}
\bibinfo{author}{Xia, C.-Y.}, \bibinfo{author}{Meloni, S.} \&
  \bibinfo{author}{Moreno, Y.}
\newblock \bibinfo{journal}{\bibinfo{title}{Effects of environment knowledge on
  agglomeration and cooperation in spatial public goods games}}.
\newblock {\emph{\JournalTitle{Adv. Complex Syst.}}}
  \textbf{\bibinfo{volume}{15}}, \bibinfo{pages}{1250056}
  (\bibinfo{year}{2012}).

\bibitem{yang_lh_csf21}
\bibinfo{author}{Yang, L.} \& \bibinfo{author}{Zhang, L.}
\newblock \bibinfo{journal}{\bibinfo{title}{Environmental feedback in spatial
  public goods game}}.
\newblock {\emph{\JournalTitle{Chaos, Solitons \& Fractals}}}
  \textbf{\bibinfo{volume}{142}}, \bibinfo{pages}{110485}
  (\bibinfo{year}{2021}).

\bibitem{esmaeili_pre18}
\bibinfo{author}{Esmaeili, S.}, \bibinfo{author}{Brown, B.~L.} \&
  \bibinfo{author}{Pleimling, M.}
\newblock \bibinfo{journal}{\bibinfo{title}{Perturbing cyclic predator-prey
  systems: How a six-species coarsening system with nontrivial in-domain
  dynamics responds to sudden changes}}.
\newblock {\emph{\JournalTitle{Phys. Rev. E}}} \textbf{\bibinfo{volume}{98}},
  \bibinfo{pages}{062105} (\bibinfo{year}{2018}).

\bibitem{shao_yx_epl19}
\bibinfo{author}{Shao, Y.}, \bibinfo{author}{Wang, X.} \& \bibinfo{author}{Fu,
  F.}
\newblock \bibinfo{journal}{\bibinfo{title}{Evolutionary dynamics of group
  cooperation with asymmetrical environmental feedback}}.
\newblock {\emph{\JournalTitle{EPL}}} \textbf{\bibinfo{volume}{126}},
  \bibinfo{pages}{40005} (\bibinfo{year}{2019}).

\bibitem{szolnoki_srep19}
\bibinfo{author}{Szolnoki, A.} \& \bibinfo{author}{Perc, M.}
\newblock \bibinfo{journal}{\bibinfo{title}{Seasonal payoff variations and the
  evolution of cooperation in social dilemmas}}.
\newblock {\emph{\JournalTitle{Sci. Rep.}}} \textbf{\bibinfo{volume}{9}},
  \bibinfo{pages}{12575} (\bibinfo{year}{2019}).

\bibitem{taitelbaum_prl20}
\bibinfo{author}{Taitelbaum, A.}, \bibinfo{author}{West, R.},
  \bibinfo{author}{Assaf, M.} \& \bibinfo{author}{Mobilia, M.}
\newblock \bibinfo{journal}{\bibinfo{title}{Population dynamics in a changing
  environment: Random versus periodic switching}}.
\newblock {\emph{\JournalTitle{Phys. Rev. Lett.}}}
  \textbf{\bibinfo{volume}{125}}, \bibinfo{pages}{048105}
  (\bibinfo{year}{2020}).

\bibitem{jansen_mb05}
\bibinfo{author}{Jansen, M. L.~A.} \emph{et~al.}
\newblock \bibinfo{journal}{\bibinfo{title}{Prolonged selection in aerobic,
  glucose-limited chemostat cultures of {\it saccharomyces cerevisiae} causes a
  partial loss of glycolytic capacity}}.
\newblock {\emph{\JournalTitle{Microbiology}}} \textbf{\bibinfo{volume}{151}},
  \bibinfo{pages}{1657--1669} (\bibinfo{year}{2005}).

\bibitem{szolnoki_epl17}
\bibinfo{author}{Szolnoki, A.} \& \bibinfo{author}{Chen, X.}
\newblock \bibinfo{journal}{\bibinfo{title}{Environmental feedback drives
  cooperation in spatial social dilemmas}}.
\newblock {\emph{\JournalTitle{EPL}}} \textbf{\bibinfo{volume}{120}},
  \bibinfo{pages}{58001} (\bibinfo{year}{2017}).

\bibitem{xie_yy_pa18}
\bibinfo{author}{Xie, Y.}, \bibinfo{author}{Chang, S.}, \bibinfo{author}{Yan,
  M.}, \bibinfo{author}{Zhang, Z.} \& \bibinfo{author}{Wang, X.}
\newblock \bibinfo{journal}{\bibinfo{title}{Environmental influences on
  cooperation in social dilemmas on networks}}.
\newblock {\emph{\JournalTitle{Physica A}}} \textbf{\bibinfo{volume}{492}},
  \bibinfo{pages}{2027--2033} (\bibinfo{year}{2018}).

\bibitem{2019-Avelino-EPL-126-68002}
\bibinfo{author}{Avelino, P.~P.} \& \bibinfo{author}{de~Oliveira, B.~F.}
\newblock \bibinfo{journal}{\bibinfo{title}{Death by starvation in
  \protect{May--Leonard} models}}.
\newblock {\emph{\JournalTitle{EPL}}} \textbf{\bibinfo{volume}{126}},
  \bibinfo{pages}{68002} (\bibinfo{year}{2019}).

\bibitem{west_jtb20}
\bibinfo{author}{West, R.} \& \bibinfo{author}{Mobilia, M.}
\newblock \bibinfo{journal}{\bibinfo{title}{Fixation properties of
  rock-paper-scissors games in fluctuating populations}}.
\newblock {\emph{\JournalTitle{J. Theor. Biol.}}}
  \textbf{\bibinfo{volume}{491}}, \bibinfo{pages}{110135}
  (\bibinfo{year}{2020}).

\bibitem{2014-Szczesny-PRE-90-032704}
\bibinfo{author}{Szczesny, B.}, \bibinfo{author}{Mobilia, M.} \&
  \bibinfo{author}{Rucklidge, A.~M.}
\newblock \bibinfo{journal}{\bibinfo{title}{Characterization of spiraling
  patterns in spatial rock-paper-scissors games}}.
\newblock {\emph{\JournalTitle{Phys. Rev. E}}} \textbf{\bibinfo{volume}{90}},
  \bibinfo{pages}{032704} (\bibinfo{year}{2014}).

\bibitem{2013-Szczesny-EPL-102-28012}
\bibinfo{author}{Szczesny, B.}, \bibinfo{author}{Mobilia, M.} \&
  \bibinfo{author}{Rucklidge, A.~M.}
\newblock \bibinfo{journal}{\bibinfo{title}{When does cyclic dominance lead to
  stable spiral waves?}}
\newblock {\emph{\JournalTitle{EPL}}} \textbf{\bibinfo{volume}{102}},
  \bibinfo{pages}{28012} (\bibinfo{year}{2013}).

\bibitem{mobilia_g16}
\bibinfo{author}{Mobilia, M.}, \bibinfo{author}{Rucklidge, A.~M.} \&
  \bibinfo{author}{Szczesny, B.}
\newblock \bibinfo{journal}{\bibinfo{title}{The influence of mobility rate on
  spiral waves in spatial rock-paper-scissors games}}.
\newblock {\emph{\JournalTitle{Games}}} \textbf{\bibinfo{volume}{7}},
  \bibinfo{pages}{24} (\bibinfo{year}{2016}).

\bibitem{frey_pa10}
\bibinfo{author}{Frey, E.}
\newblock \bibinfo{journal}{\bibinfo{title}{Evolutionary game theory:
  Theoretical concepts and applications to microbial communities}}.
\newblock {\emph{\JournalTitle{Physica A}}} \textbf{\bibinfo{volume}{389}},
  \bibinfo{pages}{4265--4298} (\bibinfo{year}{2010}).

\bibitem{szabo_pre99}
\bibinfo{author}{Szab{\'o}, G.}, \bibinfo{author}{Santos, M.~A.} \&
  \bibinfo{author}{Mendes, J. F.~F.}
\newblock \bibinfo{journal}{\bibinfo{title}{Vortex dynamics in a three-state
  model under cyclic dominance}}.
\newblock {\emph{\JournalTitle{Phys. Rev. E}}} \textbf{\bibinfo{volume}{60}},
  \bibinfo{pages}{3776--3780} (\bibinfo{year}{1999}).

\bibitem{2004-Szabo-JPAMG-31-2599}
\bibinfo{author}{Szab{\'{o}}, G.}, \bibinfo{author}{Szolnoki, A.} \&
  \bibinfo{author}{Izs{\'{a}}k, R.}
\newblock \bibinfo{journal}{\bibinfo{title}{Rock-scissors-paper game on regular
  small-world networks}}.
\newblock {\emph{\JournalTitle{Journal of Physics A: Mathematical and
  General}}} \textbf{\bibinfo{volume}{37}}, \bibinfo{pages}{2599--2609}
  (\bibinfo{year}{2004}).

\bibitem{2009-Zhang-PRE-79-062901}
\bibinfo{author}{Zhang, G.-Y.}, \bibinfo{author}{Chen, Y.},
  \bibinfo{author}{Qi, W.-K.} \& \bibinfo{author}{Qing, S.-M.}
\newblock \bibinfo{journal}{\bibinfo{title}{Four-state rock-paper-scissors
  games in constrained \protect{Newman-Watts} networks}}.
\newblock {\emph{\JournalTitle{Phys. Rev. E}}} \textbf{\bibinfo{volume}{79}},
  \bibinfo{pages}{062901} (\bibinfo{year}{2009}).

\bibitem{2014-Laird-Oikos-123-472}
\bibinfo{author}{Laird, R.~A.}
\newblock \bibinfo{journal}{\bibinfo{title}{Population interaction structure
  and the coexistence of bacterial strains playing
  ‘rock–paper–scissors’}}.
\newblock {\emph{\JournalTitle{Oikos}}} \textbf{\bibinfo{volume}{123}},
  \bibinfo{pages}{472--480} (\bibinfo{year}{2014}).

\bibitem{2014-Rulquin-PRE-89-032133}
\bibinfo{author}{Rulquin, C.} \& \bibinfo{author}{Arenzon, J.~J.}
\newblock \bibinfo{journal}{\bibinfo{title}{Globally synchronized oscillations
  in complex cyclic games}}.
\newblock {\emph{\JournalTitle{Phys. Rev. E}}} \textbf{\bibinfo{volume}{89}},
  \bibinfo{pages}{032133} (\bibinfo{year}{2014}).

\bibitem{de-oliveira_csf21}
\bibinfo{author}{\protect{de Oliveira}, B.~F.} \& \bibinfo{author}{Szolnoki,
  A.}
\newblock \bibinfo{journal}{\bibinfo{title}{Social dilemmas in off-lattice
  populations}}.
\newblock {\emph{\JournalTitle{Chaos, Solitons \& Fractals}}}
  \textbf{\bibinfo{volume}{144}}, \bibinfo{pages}{110743}
  (\bibinfo{year}{2021}).

\bibitem{2020-Bazeia-EPL-129-28002}
\bibinfo{author}{Bazeia, D.}, \bibinfo{author}{de~Moraes, M.~V.} \&
  \bibinfo{author}{de~Oliveira, B.~F.}
\newblock \bibinfo{journal}{\bibinfo{title}{Model for clustering of living
  species}}.
\newblock {\emph{\JournalTitle{EP}}} \textbf{\bibinfo{volume}{129}},
  \bibinfo{pages}{28002} (\bibinfo{year}{2020}).

\bibitem{2010-Ni-C-20-045116}
\bibinfo{author}{Ni, X.}, \bibinfo{author}{Yang, R.}, \bibinfo{author}{Wang,
  W.-X.}, \bibinfo{author}{Lai, Y.-C.} \& \bibinfo{author}{Grebogi, C.}
\newblock \bibinfo{journal}{\bibinfo{title}{Basins of coexistence and
  extinction in spatially extended ecosystems of cyclically competing
  species}}.
\newblock {\emph{\JournalTitle{Chaos}}} \textbf{\bibinfo{volume}{20}},
  \bibinfo{pages}{045116} (\bibinfo{year}{2010}).

\bibitem{2010-Ni-PRE-82-066211}
\bibinfo{author}{Ni, X.}, \bibinfo{author}{Wang, W.-X.}, \bibinfo{author}{Lai,
  Y.-C.} \& \bibinfo{author}{Grebogi, C.}
\newblock \bibinfo{journal}{\bibinfo{title}{Cyclic competition of mobile
  species on continuous space: Pattern formation and coexistence}}.
\newblock {\emph{\JournalTitle{Phys. Rev. E}}} \textbf{\bibinfo{volume}{82}},
  \bibinfo{pages}{066211} (\bibinfo{year}{2010}).

\bibitem{de-oliveira_pa21}
\bibinfo{author}{\protect{de Oliveira}, B.~F.}, \bibinfo{author}{\protect{de
  Moraes}, M.~V.}, \bibinfo{author}{Bazeia, D.} \& \bibinfo{author}{Szolnoki,
  A.}
\newblock \bibinfo{journal}{\bibinfo{title}{Mobility driven coexistence of
  living organisms}}.
\newblock {\emph{\JournalTitle{Physica A}}} \textbf{\bibinfo{volume}{572}},
  \bibinfo{pages}{125854} (\bibinfo{year}{2021}).

\bibitem{szolnoki_srep16b}
\bibinfo{author}{Szolnoki, A.} \& \bibinfo{author}{Perc, M.}
\newblock \bibinfo{journal}{\bibinfo{title}{Biodiversity in models of cyclic
  dominance is preserved by heterogeneity in site-specific invasion rates}}.
\newblock {\emph{\JournalTitle{Sci. Rep.}}} \textbf{\bibinfo{volume}{6}},
  \bibinfo{pages}{38608} (\bibinfo{year}{2016}).

\end{thebibliography}
\end{document}